\documentclass[twocolumn,letterpaper,aps,prc,superscriptaddress,showpacs,nofootinbib,floatfix]{revtex4-1}

\usepackage{graphicx}	
\usepackage{multirow}	

\usepackage[colorlinks,breaklinks]{hyperref}
\hypersetup{linkcolor=blue,citecolor=blue,filecolor=black,urlcolor=blue}

\usepackage{xspace}	

\newcommand{\pt}{\mbox{$p_T$}\xspace}
\newcommand{\pT}{\mbox{$p_T$}\xspace}

\newcommand{\mgg}{\mbox{$m_{\gamma\gamma}$}\xspace}

\newcommand{\piz}{\mbox{$\pi^0$}\xspace}

\begin{document}


\title{Azimuthal anisotropy of $\pi^0$ and $\eta$ mesons 
       in Au$+$Au collisions at $\sqrt{s_{\rm{NN}}}$ = 200 GeV}

\newcommand{\abilene}{Abilene Christian University, Abilene, Texas 79699, USA}
\newcommand{\augie}{Department of Physics, Augustana College, Sioux Falls, South Dakota 57197, USA}
\newcommand{\banaras}{Department of Physics, Banaras Hindu University, Varanasi 221005, India}
\newcommand{\barc}{Bhabha Atomic Research Centre, Bombay 400 085, India}
\newcommand{\baruch}{Baruch College, City University of New York, New York, New York, 10010 USA}
\newcommand{\bnlcoll}{Collider-Accelerator Department, Brookhaven National Laboratory, Upton, New York 11973-5000, USA}
\newcommand{\bnlphys}{Physics Department, Brookhaven National Laboratory, Upton, New York 11973-5000, USA}
\newcommand{\caucr}{University of California - Riverside, Riverside, California 92521, USA}
\newcommand{\charlesczech}{Charles University, Ovocn\'{y} trh 5, Praha 1, 116 36, Prague, Czech Republic}
\newcommand{\chonbuk}{Chonbuk National University, Jeonju, 561-756, Korea}
\newcommand{\ciae}{Science and Technology on Nuclear Data Laboratory, China Institute of Atomic Energy, Beijing 102413, P.~R.~China}
\newcommand{\cns}{Center for Nuclear Study, Graduate School of Science, University of Tokyo, 7-3-1 Hongo, Bunkyo, Tokyo 113-0033, Japan}
\newcommand{\colorado}{University of Colorado, Boulder, Colorado 80309, USA}
\newcommand{\columbia}{Columbia University, New York, New York 10027 and Nevis Laboratories, Irvington, New York 10533, USA}
\newcommand{\czechtech}{Czech Technical University, Zikova 4, 166 36 Prague 6, Czech Republic}
\newcommand{\dapnia}{Dapnia, CEA Saclay, F-91191, Gif-sur-Yvette, France}
\newcommand{\debrecen}{Debrecen University, H-4010 Debrecen, Egyetem t{\'e}r 1, Hungary}
\newcommand{\elte}{ELTE, E{\"o}tv{\"o}s Lor{\'a}nd University, H - 1117 Budapest, P{\'a}zm{\'a}ny P. s. 1/A, Hungary}
\newcommand{\ewha}{Ewha Womans University, Seoul 120-750, Korea}
\newcommand{\fit}{Florida Institute of Technology, Melbourne, Florida 32901, USA}
\newcommand{\fsu}{Florida State University, Tallahassee, Florida 32306, USA}
\newcommand{\gsu}{Georgia State University, Atlanta, Georgia 30303, USA}
\newcommand{\hiroshima}{Hiroshima University, Kagamiyama, Higashi-Hiroshima 739-8526, Japan}
\newcommand{\ihepprot}{IHEP Protvino, State Research Center of Russian Federation, Institute for High Energy Physics, Protvino, 142281, Russia}
\newcommand{\illuiuc}{University of Illinois at Urbana-Champaign, Urbana, Illinois 61801, USA}
\newcommand{\inrras}{Institute for Nuclear Research of the Russian Academy of Sciences, prospekt 60-letiya Oktyabrya 7a, Moscow 117312, Russia}
\newcommand{\instpasczech}{Institute of Physics, Academy of Sciences of the Czech Republic, Na Slovance 2, 182 21 Prague 8, Czech Republic}
\newcommand{\isu}{Iowa State University, Ames, Iowa 50011, USA}
\newcommand{\jaea}{Advanced Science Research Center, Japan Atomic Energy Agency, 2-4 Shirakata Shirane, Tokai-mura, Naka-gun, Ibaraki-ken 319-1195, Japan}
\newcommand{\jinrdubna}{Joint Institute for Nuclear Research, 141980 Dubna, Moscow Region, Russia}
\newcommand{\jyvaskyla}{Helsinki Institute of Physics and University of Jyv{\"a}skyl{\"a}, P.O.Box 35, FI-40014 Jyv{\"a}skyl{\"a}, Finland}
\newcommand{\kek}{KEK, High Energy Accelerator Research Organization, Tsukuba, Ibaraki 305-0801, Japan}
\newcommand{\korea}{Korea University, Seoul, 136-701, Korea}
\newcommand{\kurchatov}{Russian Research Center ``Kurchatov Institute", Moscow, 123098 Russia}
\newcommand{\kyoto}{Kyoto University, Kyoto 606-8502, Japan}
\newcommand{\labllr}{Laboratoire Leprince-Ringuet, Ecole Polytechnique, CNRS-IN2P3, Route de Saclay, F-91128, Palaiseau, France}
\newcommand{\lahorelums}{Physics Department, Lahore University of Management Sciences, Lahore, Pakistan}
\newcommand{\lawllnl}{Lawrence Livermore National Laboratory, Livermore, California 94550, USA}
\newcommand{\losalamos}{Los Alamos National Laboratory, Los Alamos, New Mexico 87545, USA}
\newcommand{\lpc}{LPC, Universit{\'e} Blaise Pascal, CNRS-IN2P3, Clermont-Fd, 63177 Aubiere Cedex, France}
\newcommand{\lund}{Department of Physics, Lund University, Box 118, SE-221 00 Lund, Sweden}
\newcommand{\maryland}{University of Maryland, College Park, Maryland 20742, USA}
\newcommand{\mass}{Department of Physics, University of Massachusetts, Amherst, Massachusetts 01003-9337, USA }
\newcommand{\michigan}{Department of Physics, University of Michigan, Ann Arbor, Michigan 48109-1040, USA}
\newcommand{\muenster}{Institut fur Kernphysik, University of Muenster, D-48149 Muenster, Germany}
\newcommand{\muhlenberg}{Muhlenberg College, Allentown, Pennsylvania 18104-5586, USA}
\newcommand{\myongji}{Myongji University, Yongin, Kyonggido 449-728, Korea}
\newcommand{\nagasaki}{Nagasaki Institute of Applied Science, Nagasaki-shi, Nagasaki 851-0193, Japan}
\newcommand{\newmex}{University of New Mexico, Albuquerque, New Mexico 87131, USA }
\newcommand{\nmsu}{New Mexico State University, Las Cruces, New Mexico 88003, USA}
\newcommand{\ohio}{Department of Physics and Astronomy, Ohio University, Athens, Ohio 45701, USA}
\newcommand{\ornl}{Oak Ridge National Laboratory, Oak Ridge, Tennessee 37831, USA}
\newcommand{\orsay}{IPN-Orsay, Universite Paris Sud, CNRS-IN2P3, BP1, F-91406, Orsay, France}
\newcommand{\peking}{Peking University, Beijing 100871, P.~R.~China}
\newcommand{\pnpi}{PNPI, Petersburg Nuclear Physics Institute, Gatchina, Leningrad region, 188300, Russia}
\newcommand{\riken}{RIKEN Nishina Center for Accelerator-Based Science, Wako, Saitama 351-0198, Japan}
\newcommand{\rikjrbrc}{RIKEN BNL Research Center, Brookhaven National Laboratory, Upton, New York 11973-5000, USA}
\newcommand{\rikkyo}{Physics Department, Rikkyo University, 3-34-1 Nishi-Ikebukuro, Toshima, Tokyo 171-8501, Japan}
\newcommand{\saispbstu}{Saint Petersburg State Polytechnic University, St. Petersburg, 195251 Russia}
\newcommand{\saopaulo}{Universidade de S{\~a}o Paulo, Instituto de F\'{\i}sica, Caixa Postal 66318, S{\~a}o Paulo CEP05315-970, Brazil}
\newcommand{\seoulnat}{Seoul National University, Seoul, Korea}
\newcommand{\stonybrkc}{Chemistry Department, Stony Brook University, SUNY, Stony Brook, New York 11794-3400, USA}
\newcommand{\stonycrkp}{Department of Physics and Astronomy, Stony Brook University, SUNY, Stony Brook, New York 11794-3400, USA}
\newcommand{\tenn}{University of Tennessee, Knoxville, Tennessee 37996, USA}
\newcommand{\titech}{Department of Physics, Tokyo Institute of Technology, Oh-okayama, Meguro, Tokyo 152-8551, Japan}
\newcommand{\tsukuba}{Institute of Physics, University of Tsukuba, Tsukuba, Ibaraki 305, Japan}
\newcommand{\vandy}{Vanderbilt University, Nashville, Tennessee 37235, USA}
\newcommand{\waseda}{Waseda University, Advanced Research Institute for Science and Engineering, 17 Kikui-cho, Shinjuku-ku, Tokyo 162-0044, Japan}
\newcommand{\weizmann}{Weizmann Institute, Rehovot 76100, Israel}
\newcommand{\wigner}{Institute for Particle and Nuclear Physics, Wigner Research Centre for Physics, Hungarian Academy of Sciences (Wigner RCP, RMKI) H-1525 Budapest 114, POBox 49, Budapest, Hungary}
\newcommand{\yonsei}{Yonsei University, IPAP, Seoul 120-749, Korea}
\affiliation{\abilene}
\affiliation{\augie}
\affiliation{\banaras}
\affiliation{\barc}
\affiliation{\baruch}
\affiliation{\bnlcoll}
\affiliation{\bnlphys}
\affiliation{\caucr}
\affiliation{\charlesczech}
\affiliation{\chonbuk}
\affiliation{\ciae}
\affiliation{\cns}
\affiliation{\colorado}
\affiliation{\columbia}
\affiliation{\czechtech}
\affiliation{\dapnia}
\affiliation{\debrecen}
\affiliation{\elte}
\affiliation{\ewha}
\affiliation{\fit}
\affiliation{\fsu}
\affiliation{\gsu}
\affiliation{\hiroshima}
\affiliation{\ihepprot}
\affiliation{\illuiuc}
\affiliation{\inrras}
\affiliation{\instpasczech}
\affiliation{\isu}
\affiliation{\jaea}
\affiliation{\jinrdubna}
\affiliation{\jyvaskyla}
\affiliation{\kek}
\affiliation{\korea}
\affiliation{\kurchatov}
\affiliation{\kyoto}
\affiliation{\labllr}
\affiliation{\lahorelums}
\affiliation{\lawllnl}
\affiliation{\losalamos}
\affiliation{\lpc}
\affiliation{\lund}
\affiliation{\maryland}
\affiliation{\mass}
\affiliation{\michigan}
\affiliation{\muenster}
\affiliation{\muhlenberg}
\affiliation{\myongji}
\affiliation{\nagasaki}
\affiliation{\newmex}
\affiliation{\nmsu}
\affiliation{\ohio}
\affiliation{\ornl}
\affiliation{\orsay}
\affiliation{\peking}
\affiliation{\pnpi}
\affiliation{\riken}
\affiliation{\rikjrbrc}
\affiliation{\rikkyo}
\affiliation{\saispbstu}
\affiliation{\saopaulo}
\affiliation{\seoulnat}
\affiliation{\stonybrkc}
\affiliation{\stonycrkp}
\affiliation{\tenn}
\affiliation{\titech}
\affiliation{\tsukuba}
\affiliation{\vandy}
\affiliation{\waseda}
\affiliation{\weizmann}
\affiliation{\wigner}
\affiliation{\yonsei}
\author{A.~Adare} \affiliation{\colorado}
\author{S.~Afanasiev} \affiliation{\jinrdubna}
\author{C.~Aidala} \affiliation{\mass} \affiliation{\michigan}
\author{N.N.~Ajitanand} \affiliation{\stonybrkc}
\author{Y.~Akiba} \affiliation{\riken} \affiliation{\rikjrbrc}
\author{H.~Al-Bataineh} \affiliation{\nmsu}
\author{J.~Alexander} \affiliation{\stonybrkc}
\author{K.~Aoki} \affiliation{\kyoto} \affiliation{\riken}
\author{Y.~Aramaki} \affiliation{\cns}
\author{E.T.~Atomssa} \affiliation{\labllr}
% ...
\collaboration{PHENIX Collaboration} \noaffiliation

\date{\today}

\begin{abstract}
The azimuthal anisotropy coefficients $v_2$ and $v_4$ of $\pi^0$ and 
$\eta$ mesons are measured in Au$+$Au collisions at $\sqrt{s_{\rm NN}} = 
200$~GeV as a function of transverse momentum $p_T$ (1--14 GeV/$c$) and 
centrality. The extracted $v_2$ coefficients are found to be consistent 
between the two meson species over the measured $p_T$ range.  The ratio 
of $v_4/v_2^2$ for $\pi^0$ mesons is found to be independent of $p_T$ 
for 1--9 GeV/$c$, implying a lack of sensitivity of the ratio to the 
change of underlying physics with $p_T$. Furthermore, the ratio of 
$v_4/v_2^2$ is systematically larger in central collisions, which may 
reflect the combined effects of fluctuations in the initial collision 
geometry and finite viscosity in the evolving medium.
\end{abstract}

	

\maketitle

\section{Introduction}

A novel form of nuclear matter, where quarks and gluons are deconfined 
yet interact strongly with each other, is produced in heavy ion 
collisions at the relativistic heavy ion collider 
(RHIC)~\cite{Adcox:2004mh,Adams:2005dq,Arsene:2004fa,Back:2004je} and 
the large hadron collider (LHC)~\cite{Muller:2012zq}. The hydrodynamic 
expansion of this matter, as well as its interactions with hard 
scattered partons, result in the anisotropic emission of 
hadrons~\cite{Heinz:2001xi,Gyulassy:2000gk}. Measurements of azimuthal 
anisotropy for particle production provide valuable information on the 
transport properties of the 
matter~\cite{Adare:2006nq,Lacey:2006bc,Romatschke:2007mq}.

The magnitude of the anisotropy can be studied from the azimuthal angle 
($\phi$) distribution of particles relative to the second 
order\footnote{The $v_{4}$ coefficient can be measured with respect to the 
second order event plane or the fourth order event plane.  In the analysis 
presented in this work, all $v_{4}$ coefficients are measured with respect 
to the second order event plane.} event plane (EP) angle ($\Phi$) 
accumulated over many events~\cite{Poskanzer:1998yz,Afanasiev:2009wq}:
\begin{equation}
\label{eq:flow}
\frac{dN}{d\Delta\phi}\propto1+2\sum_{k=1}^{\infty}v_{2k}\cos ( 2k\Delta\phi\ );, \\\label{eq:flow2}
\end{equation}
where $\Delta\phi=(\phi-\Phi)$ and $v_{2k}$ are even-order Fourier 
coefficients, which generally are nonzero around the elliptic flow 
EP~\cite{Poskanzer:1998yz}. In the {\em event-plane} 
method, an 
estimated EP angle $\Psi$ is determined from the particles in 
the event.  Due to the finite number of particles used to determine $\Psi$, 
the $\Psi$ angle is an approximation of the true EP angle $\Phi$. The 
coefficient $v_{2k}$ is measured by correlating particles with $\Psi$ to 
obtain the raw values 
$v_{2k}^{{\rm obs}}=\langle \cos ( 2k [ \phi-\Psi ] ) \rangle$, which are 
then corrected by a 
resolution factor (${\rm Res}\{2k\Psi\}$) that accounts for the spread 
of $\Psi$ about $\Phi$~\cite{Poskanzer:1998yz}:

\begin{eqnarray} 
\label{eq:corr} 
v_{2k}=\frac{v_{2k}^{\rm obs}}{{\rm Res}\{2k\Psi\}}
=\frac{\langle \cos ( 2k [ \phi-\Psi ] ) \rangle}
      {\langle \cos ( 2k [ \Psi-\Phi ] ) \rangle}\;\;  k=1, 2.
\end{eqnarray}
To minimize the nonflow biases from dijets, the particles used 
to estimate $\Psi$ are selected in a pseudorapidity range that is well 
separated (typically one unit or more) from the particles used to 
evaluate $v_{2k}^{{\rm obs}}$~\cite{Adare:2008ae}.

Recently, experiments at RHIC and LHC have measured significant $v_{2k}$ 
values for $k=1$--3 for various particle species and over a broad range in 
$\pT$~\cite{Adams:2003zg,Adare:2010ux,ATLAS:2012at,Abelev:2012di,Chatrchyan:2012ta}. 
For particles with low transverse momenta ($\pT\alt3$~GeV/$c$), the 
coefficients are understood in terms of pressure-driven {\it flow} in an 
initial ``almond-shaped" collision zone produced in noncentral 
collisions \cite{Ollitrault:1992bk}. For higher transverse momenta 
($\pT\agt6$--10 GeV/$c$), the anisotropy reflected by the $v_{2k}$ 
coefficients can be attributed to jet quenching \cite{Gyulassy:1993hr} 
-- the process by which hard scattered partons interact and lose energy 
in the hot and dense medium prior to fragmenting into hadrons. This 
energy loss manifests as a suppression of hadron 
yields~\cite{Adcox:2001jp}, which depends on the average path-length 
that partons propagate through the medium~\cite{Adare:2010sp,Adare:2012wg}, 
and $v_{2}$ for example stems from the fact that the partons traveling 
in a direction parallel to the $\Phi$ angle are less suppressed than 
those traveling in the direction perpendicular to the $\Phi$ 
angle~\cite{Gyulassy:2000gk}.

The present work exploits various PHENIX detector subsystems with a 
broad range in pseudorapidity for the EP determination, and provides 
detailed differential measurements of $v_{2}$ and $v_4$ for $\pi^0$ 
mesons and $v_2$ for $\eta$ mesons in $\sqrt{s_{\rm NN}} = 200$~GeV 
Au$+$Au collisions. The $v_2$ and $v_4$ measurements for $\pi^0$ mesons 
extend our earlier work~\cite{Adare:2010sp,Adare:2012wg}. The $v_2$ 
measurements for $\eta$ mesons probe the particle species dependence of 
jet-quenching and test the consistency of the data with medium-induced 
partonic energy loss prior to vacuum 
hadronization~\cite{Adare:2012vq,Abelev:2012di}. In this vacuum 
hadronization picture, high $\pT$ $\pi^0$ and $\eta$ mesons are thought 
to arise from fragmentation of energetic partons after they lose energy 
in the medium, and hence the two types of mesons are expected to show 
similar level of suppression~\cite{Adler:2006hu} and similar path-length 
dependence or $v_2$. Furthermore, this analysis provides a test of the 
previous observed scaling relation between $v_2$ and $v_4$, i.e. the 
observation that $v_4/v_2^2$ ratio is approximately independent of 
$\pT$~\cite{Adams:2003zg,Adare:2010ux,Lacey:2011ug,ATLAS:2012at}. This 
analysis also allows a detailed study of the biases from dijets in the 
determination of the event plane.

\section{Measurement}
\subsection{Data set and Centrality}

The measurements are based on a Au$+$Au collision data set at 
$\sqrt{s_{\rm NN}}=200$ GeV collected during the 2007 running period. 
The minimum bias events are selected by the beam-beam counters (BBC). 
The collision vertex along the beam direction, $z$, is measured by the 
BBC. After an offline vertex cut of $|z|<30$~cm and run quality 
selections, a total of $\sim3.5\times 10^9$ minimum bias events are 
obtained. Event centrality for these events are determined by the number 
of charged particles detected in the BBCs~\cite{Adare:2008ae}. A Glauber 
model Monte-Carlo simulation~\cite{Adler:2003au} that includes the 
responses of the BBC is used to estimate, for each centrality selection, 
the average number of participating nucleons $N_{\rm part}$.

\subsection{Event plane measurement}

The EP angle is estimated using several detectors installed 
symmetrically on both sides of the nominal collision point along the 
beamline: the BBC~\cite{Afanasiev:2009aa}, the muon piston calorimeters 
(MPC)~\cite{Adare:2011sc}, and the reaction-plane detectors 
(RXN)~\cite{Richardson:2010hm}. The BBCs comprise two sets of 64 
\v{C}erenkov counter modules, located at $z=\pm144$ cm from the nominal 
collision point and measure the number of charged particles over the 
pseudorapidity region $3.1 < |\eta| < 3.9$. Each MPC is equipped with 
PbWO$_4$ crystal scintillator towers. The north MPC has 220 towers 
spanning $3.1<\eta<3.9$, while the south MPC has 196 towers spanning 
$-3.7<\eta<-3.1$. The MPCs have almost the same azimuth and $\eta$ 
coverage as the BBCs, but have finer granularity and detect both charged 
and neutral particles, and hence have better EP resolution. The RXNs are 
situated at $z=\pm40$~cm from the nominal interaction point. Each 
comprises 12 azimuthally segmented scintillator paddles with 
photomultiplier read out. They are covered with a 2~cm (3.6 radiation 
length) thick lead photon converter and are sensitive to both charged 
particles and photons. The RXNs cover the pseudorapidity region $1.0 < 
|\eta| < 2.8$. They are further subdivided into an outer part (RXN$_{\rm 
out}$, $1.0 < |\eta| < 1.5$) and an inner part (RXN$_{\rm in}$, 
$1.5 < |\eta| < 2.8$).

Table~\ref{tab:2} outlines the $\eta$ acceptance of BBCs, MPCs and RXNs, 
as well as several combined detectors from which the EP are estimated. 
These combinations allow for a reliable estimate of the systematic 
uncertainties in this measurement. The results reported in this paper 
use the EP from MPC+RXN$_{\rm in}$~\cite{Adare:2010sp}, which provides 
very good resolution and minimizes the possible nonflow biases from jets 
and dijets~\cite{Adare:2008ae}.

\begin{table}
\caption{\label{tab:2} Summary of the $\eta$ coverage for the detector 
combinations used for the event plane measurements.}
\begin{ruledtabular} \begin{tabular}{cc}
Detectors                             &       $\eta$ range       \\ \hline
BBC                 &$\pm[3.1,3.9]$             \\
MPC                 &$[-3.7,-3.1]$,$[3.1,3.9]$  \\
${\rm RXN_{\rm in}}$     &$\pm[1.5,2.8]$           \\
${\rm RXN_{\rm out}}$    &$\pm[1.0,1.5]$           \\
RXN(in + out)       &$\pm[1.0,2.8]$           \\
$\rm MPC+RXN_{\rm in}$   & $\pm[1.5,2.8]$, $[-3.7,-3.1]$,$[3.1,3.9]$\\
\end{tabular}  \end{ruledtabular}
\end{table}

The resolution factor ${\rm Res}\{2k\Psi\}$ is determined using the two 
subevents (2SE) and three subevents (3SE) 
methods~\cite{Poskanzer:1998yz}, as outlined in our previous 
analyses~\cite{Adare:2010ux,Adare:2012vq}. In the 2SE method, the signal 
of a given detector combination in Table~\ref{tab:2} is divided into two 
subevents covering equal pseudorapidity ranges in opposite hemispheres. 
The resolution of each subevent is calculated directly from the 
correlation between the two subevents:
\begin{eqnarray}
\label{eq:mep3}
{{\rm Res}\{2k\Psi^{{\rm A}}\}}= 
{{\rm Res}\{2k\Psi^{{\rm B}}\}}= 
\sqrt{\langle {\cos ( 2k [ \Psi^{\mathrm A}-\Psi^{\mathrm B} ] ) }\rangle}.
\end{eqnarray}
The resolution can be expressed analytically as:
\begin{eqnarray}
\label{eq:mep2}
\nonumber
{{\rm Res}\{2k\Psi\}}&=&
\langle {\cos ( 2k\Delta \Psi ) } \rangle \\ &=&
\frac{{\chi\sqrt \pi }}{2} e^ {- \frac{{\chi^2 }}{2}}\left[ {I_{\frac{k - 1}{2}} 
(\frac{{\chi ^2 }}{2}) + I_{\frac{k + 1}{2}} (\frac{{\chi^2 }}{2})} \right],
\end{eqnarray}
where $I_{\alpha}$ are modified Bessel functions of the first kind, and 
the resolution parameter $\chi\propto \sqrt{M}$ is related to the 
multiplicity $M$. The resolution parameter of the full detector is 
determined as $\chi=\sqrt{2}\chi_{{\rm A}}=\sqrt{2}\chi_{{\rm B}}$, 
which is then used to determine ${{\rm Res}\{2k\Psi\}}$ via 
Eq.~\ref{eq:mep2}.

The 3SE method determines the resolution factor of a given detector from 
the correlations of its EP with those for two other detectors in 
different pseudorapidity ranges:
\begin{eqnarray}
\nonumber
{\rm Res}\{2k\Psi^A\}= 
\sqrt{\frac{\langle {\cos ( 2k [\Psi^A-\Psi^B ] )} \rangle 
            \langle {\cos ( 2k [\Psi^A-\Psi^C ] )} \rangle}
           {\langle {\cos ( 2k [\Psi^B-\Psi^C ] )} \rangle}}.\\\label{eq:mm3}
\end{eqnarray}\normalsize
The main advantage of the 3SE method is that for a given detector $A$, 
there are many choices of detectors $B$ and $C$, which provide 
independent estimates of the resolution of $A$. The differences between 
the resolution estimates for the 2SE and 3SE methods are included in the 
evaluation of systematic uncertainties.

\begin{figure}
\includegraphics[width=1.0\linewidth]{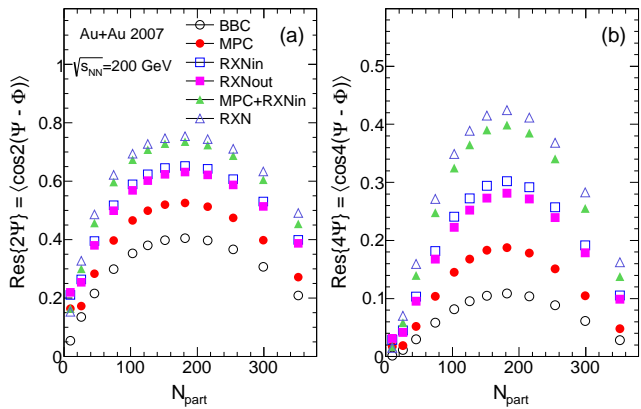}
\caption{\label{fig:rp4} (Color online) Resolutions for $v_2$ (left) and 
$v_4$ (right) calculated using various detectors for event plane.}
\end{figure}

The left panel of Fig.~\ref{fig:rp4} summarizes the $N_{\rm part}$ 
dependence of the 2SE resolution factors for various detector 
combinations as indicated. The resolution factors peak around $N_{\rm 
part}\sim180$ (i.e. the 20\%--30\% centrality bin) with maximum values 
of 0.75 for RXN, 0.53 for MPC and 0.4 for the BBC. The resolution 
factors for RXN$_{\rm in}$ and RXN$_{\rm out}$ are similar and show a 
maximum of $\sim0.65$. The resolution factors for MPC+RXN$_{\rm in}$ 
are very close to those for the full RXN. The right panel of 
Fig.~\ref{fig:rp4} shows resolution factors for $v_4$ that are much 
smaller than those for $v_2$. The associated resolution factors also 
peak for $N_{\rm part}\sim180$, reaching maximum values of 0.45, 0.18, 
0.1 and 0.4 for the full RXN, MPC, BBC and MPC+RXN$_{\rm in}$ 
respectively.

\subsection{Measurements of $\pi^0$ and $\eta$ meson anisotropy}

\subsubsection{Reconstruction of $\pi^0$ and $\eta$ mesons}

Neutral pion and $\eta$ mesons are measured via their $\gamma\gamma$ 
decay channel in the electromagnetic calorimeter (EMCal, $|\eta|<0.35$) 
~\cite{nimemc}. The EMCal comprises the 
lead-scintillator and lead-glass subdetectors, covering 0.75$\pi$ and 
0.25$\pi$ in azimuth, respectively.  Photons are identified using various 
cuts on the shower shape observed in the EMCal, as well as by comparing 
the observed shapes to a template profile function measured from test beam 
data. The invariant mass \mgg is calculated for photon pairs, which pass 
an energy asymmetry cut $\alpha=|E_1-E_2|/(E_1+E_2)<0.8$ and have a 
minimum separation of 8\,cm between their impact points in the EMCal. The 
combinatorial background distribution in \mgg is estimated with the event 
mixing technique where the two photons are selected from different events 
satisfying similar global requirements such as vertex, centrality and 
event plane direction. The mixed-event \mgg distributions are then 
normalized in the sidebands of the \piz and $\eta$ peaks in the real event 
distributions and subtracted. A small residual background is parameterized 
by a first-order polynomial in the regions below and above the \piz and 
$\eta$ peak, and then subtracted from the \mgg distribution. The raw \piz 
and $\eta$ meson yields are calculated by integrating a $\pm$2$\sigma$ 
window in \mgg around their respective peaks. This window is varied 
($\pm2\sigma$ vs $\pm3\sigma$) to check the stability of the yield.  The 
ratios of signal to background (S/N) for \piz and $\eta$ mesons varies 
strongly with $\pT$ and centrality. The S/N value generally increases with 
$\pT$ and increases from central to peripheral collisions. The values of 
S/N for \piz and $\eta$ mesons are given in Table~\ref{tab:sn} for events 
in 0\%--20\% collisions. In this analysis, the reconstruction of the \piz 
and 
$\eta$ mesons is limited to $\pT>1$ GeV/$c$ and $\pT>3$ GeV/$c$ 
respectively, where the yields of the two mesons can be extracted with 
relatively small uncertainty. Further details can be found in 
Ref.~\cite{Adare:2012wg,Adare:2010dc}.

\begin{table}
\caption{\label{tab:sn} The ratios of signal to background (S/N) for \piz 
and $\eta$ mesons at several $\pT$ in 0\%--20\% most central collisions. 
The values are given for pairs integrated in a $\pm$2$\sigma$ window in 
\mgg around their respective peaks.}
\begin{ruledtabular} \begin{tabular}{cccccc}
  $\pT$ (GeV/$c$)    & 1    & 3      & 6   & 10&\\ \hline
\piz S/N             & 0.01   & 0.15    & 2       & 8& \\
eta S/N              & N/A  & 0.01     & 0.1     & 1 &\\
\end{tabular} \end{ruledtabular}
\end{table}

\subsubsection{The $dN/d\Delta\phi$ method}

The first method for the extraction of $v_{2}$ and $v_4$ follows the 
analysis method outlined in our prior work~\cite{Afanasiev:2009aa}. The 
photon pairs in each $\pt$ and centrality bin are divided according to 
their angle relative to the estimated EP angle, $\Delta\phi=\phi-\Psi$, 
into six bins in the interval of [$0,\pi/2$]. The yields of $\pi^0$ and 
$\eta$ mesons are extracted independently in each bin and then 
parameterized by:
\begin{eqnarray}
\frac{dN}{d\Delta\phi} = 
N_{0}[1+2v_{2}^{\rm obs}\cos(2\Delta\phi)+2v_{4}^{\rm obs}\cos(4\Delta\phi)], 
\end{eqnarray}
to obtain $v_{2k}^{\rm obs}$. The values of $v_{2k}^{\rm obs}$ are also 
calculated directly via a discrete Fourier transform:
\begin{eqnarray}
v_{2k}^{\rm obs}=\frac{\sum_{i=1}^{6} N_i\cos(2k\Delta\phi_i)}{\sum_{i=1}^{6} N_i},
\end{eqnarray}
where the $N_i$ stands for the yield in the $i^{\rm th}$ angular bin. 
The two results are found to be consistent within 2\% of their central 
values. Because of the finite bin-width in $\Delta\phi$, the extracted 
$v_{2k}^{\rm obs}$ values need to be corrected up by a smearing factor 
\mbox{$\sigma_k= \frac{k \delta}{\sin(k \delta)}$}, which accounts for 
the finite bin width $\delta=\pi/12$.

The $v_2$ and $v_4$ values for this method are obtained by applying both 
the resolution correction (Eq.~\ref{eq:corr}) and smearing correction to 
$v_{2}^{\rm obs}$ and $v_{4}^{\rm obs}$ for each centrality and $\pT$ 
selection. To check the sensitivity of the yield extraction on our 
choices of bin width in $\Delta\phi$, the $v_2$ and $v_4$ values are 
also calculated using 18 bins in $\Delta\phi$ for $p_T<10$ GeV/$c$. The 
results are found to be consistent with the 6 bin results within 3\% of 
the average of the two measurements.

\subsubsection{Invariant mass method}

The second method of extracting $v_{2}$ and $v_{4}$ for $\pi^0$ mesons  
follows 
the procedure outlined in 
Refs.~\cite{Borghini:2004ra,Abelev:2008ae,Afanasiev:2007tv}. In this 
method, the anisotropy of same-event or foreground (frg) pairs and 
mixed-events or background (bkg) pairs are determined as a function of 
$m_{\gamma\gamma}$, denoted as $v_{2k}^{\rm frg}(m_{\gamma\gamma})$ and 
$v_{2k}^{\rm bkg}(m_{\gamma\gamma})$, respectively. The anisotropy of 
foreground pairs can be expressed as the sum of the contributions from 
the signal pairs (sig) and the background pairs in each 
$m_{\gamma\gamma}$ bin:
\begin{eqnarray}
N_{\rm frg} \, v_{2k}^{\rm frg} &=& N_{\rm sig} \, v_{2k}^{\rm sig}
+N_{\rm bkg} \, v_{2k}^{\rm bkg}\\
N_{\rm frg} &=& N_{\rm sig}+N_{\rm bkg}\;,
\end{eqnarray}
which gives the expression:
\begin{eqnarray}
\nonumber
v_{2k}^{\rm sig} (m_{\gamma\gamma}) &=& \frac{v_{2k}^{\rm frg} 
(m_{\gamma\gamma})-v_{2k}^{\rm bkg}(m_{\gamma\gamma})
[1-R(m_{\gamma\gamma})]}{R(m_{\gamma\gamma})},\\\label{eq:inv}
\end{eqnarray}
where $R= N_{\rm sig}/(N_{\rm sig}+N_{\rm bkg})$ is the fraction of the 
total number of pairs comprising the signal.

\begin{figure}[th]
\begin{center}
\includegraphics[width=1.0\linewidth]{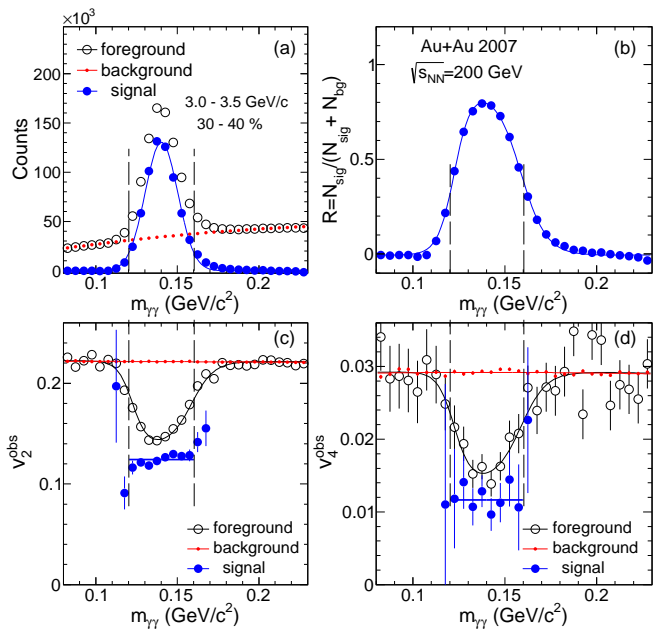}
\end{center}
\caption{\label{fig:pi0m5} (Color online) The $m_{\gamma\gamma}$ 
distributions used to obtain $v_2^{{\rm obs}}$ and $v_4^{{\rm obs}}$ 
values. (a) Distributions of foreground $N_{\rm frg}$, background 
$N_{\rm bkg}$ and signal $N_{\rm sig}$, (b) the signal fraction $R= 
N_{\rm sig}/(N_{\rm sig}+N_{\rm bkg})$, (c) the $v_2^{{\rm obs}}$ 
distributions for foreground, background and signal, and (d) the 
$v_4^{{\rm obs}}$ distributions for foreground, background and signal. 
The vertical dashed lines indicate the two standard deviation window 
around the $\pi^0$ peak.}
\end{figure}

Figure~\ref{fig:pi0m5} illustrates the steps in calculating $v_{2k}^{\rm 
sig}(m_{\gamma\gamma})$ for a given $p_T$ and centrality bin. 
Figure~\ref{fig:pi0m5}(a) shows the distributions of the foreground 
$N_{\rm frg}(m_{\gamma\gamma})$, total background $N_{\rm 
bkg}(m_{\gamma\gamma})$, and the extracted signal $N_{\rm 
sig}(m_{\gamma\gamma})$. The $N_{\rm bkg}$ is determined from 
mixed-events in concert with a linear parameterization of the residual 
background, as described earlier. Figure~\ref{fig:pi0m5}(b) shows the 
resulting $R(m_{\gamma\gamma})$. The signal anisotropy coefficients are 
calculated directly as $v_{2k}^{\rm 
sig}=\langle\cos(2k(\phi-\Psi))\rangle$ as a function of 
$m_{\gamma\gamma}$, as shown by the open circles in 
Figs.~\ref{fig:pi0m5}(c)-(d). They have a concave shape in the $\pi^0$ 
signal region, and show a minimum at the $\pi^0$ mass peak. In regions 
far away from the $\pi^0$ mass, the $v_{2k}^{\rm sig}$ values vary 
slowly with $m_{\gamma\gamma}$, reflecting the anisotropy of the 
background. The concave shape is a general feature of the invariant mass 
method for reconstructing $v_2$ of decay 
particles~\cite{Borghini:2004ra,Abelev:2008ae,Afanasiev:2007tv}: the 
background photon pairs on average have a small opening angle owing to the 
asymmetry cut, and hence they have a larger anisotropy compared to 
photons from $\pi^0$ decay.

The two photons of the mixed-event pairs are chosen from an event class 
with similar event plane orientations, so they have a sizable anisotropy 
$v_{2k}^{\rm bg,mix}(m_{\gamma\gamma})$. However, because the two events 
used to construct the mixed-event do not have exactly the same EP angle, 
$v_{2k}^{{\rm bg,mix}}(m_{\gamma\gamma})$ is smaller than the 
$v_{2k}^{\rm bkg}(m_{\gamma\gamma})$ by a factor that depends on the EP 
resolution and the bin width of the EP class used for event mixing, but 
is independent of $m_{\gamma\gamma}$. Hence the $v_{2k}^{\rm 
bkg}(m_{\gamma\gamma})$ value is obtained by first scaling the measured 
$v_{2k}^{\mathrm {bg,mix}}(m_{\gamma\gamma})$ distribution to match the 
$v_{2k}^{\rm frg}(m_{\gamma\gamma})$ in regions three standard 
deviations away from the $\pi^0$ peak. The resulting $v_{2k}^{\rm 
bkg}(m_{\gamma\gamma})$ distributions are indicated by the small dotted 
symbols in Figs.~\ref{fig:pi0m5} (c)-(d). The anisotropy coefficients of 
the signals $v_{2k}^{\rm sig}$ are then calculated bin-by-bin in 
$m_{\gamma\gamma}$ according to Eq.~\ref{eq:inv}. They are shown by the 
solid symbols in Figs.~\ref{fig:pi0m5}(c)-(d).

Figure~\ref{fig:pi0m5}(c) shows a slight increase of $v_{2}^{\rm sig}$ 
at the upper end of the $\pi^0$ peak. This increase is due to 
overlapping clusters, which also manifests as an excess in the Gaussian 
fit to the signal Figs.~\ref{fig:pi0m5}(a). Since the $v_{2}^{\rm sig}$ 
measurement is dominated by the points in the vicinity of the $\pi^0$ 
peak, the influence of this increase on $v_{2}^{\rm sig}$ is less than 
3\% of its magnitude and it is included in the systematic uncertainties.

\begin{figure}[th]
\begin{center}
\includegraphics[width=1.0\linewidth]{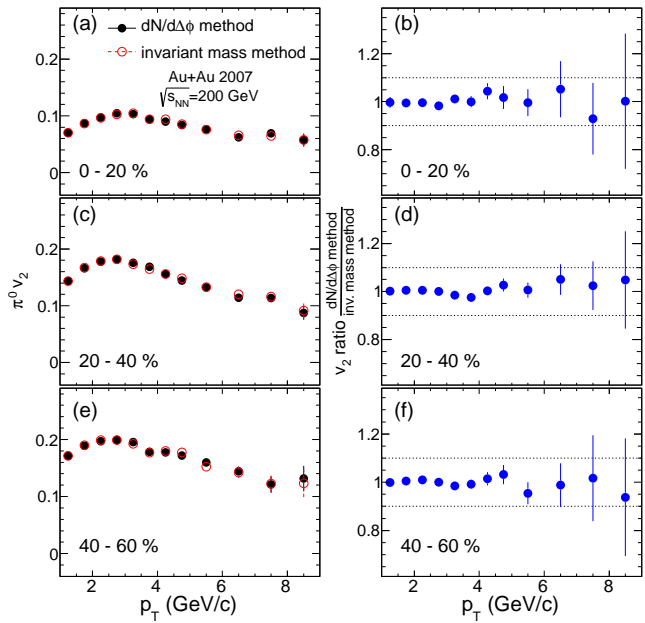}
\end{center}
\caption{\label{fig:pi0m6} (Color online) Comparison of $\pi^0$ $v_2$ 
for the $dN/d\Delta\phi$ and the invariant mass methods of analysis for 
several centrality selections (left panels). The corresponding ratios 
are shown in the right panels, with the dashed curves indicate a $\pm 
10$\% deviation from unity.}
\end{figure} %
\begin{figure}[ht] %
\begin{center}
\includegraphics[width=1.0\linewidth]{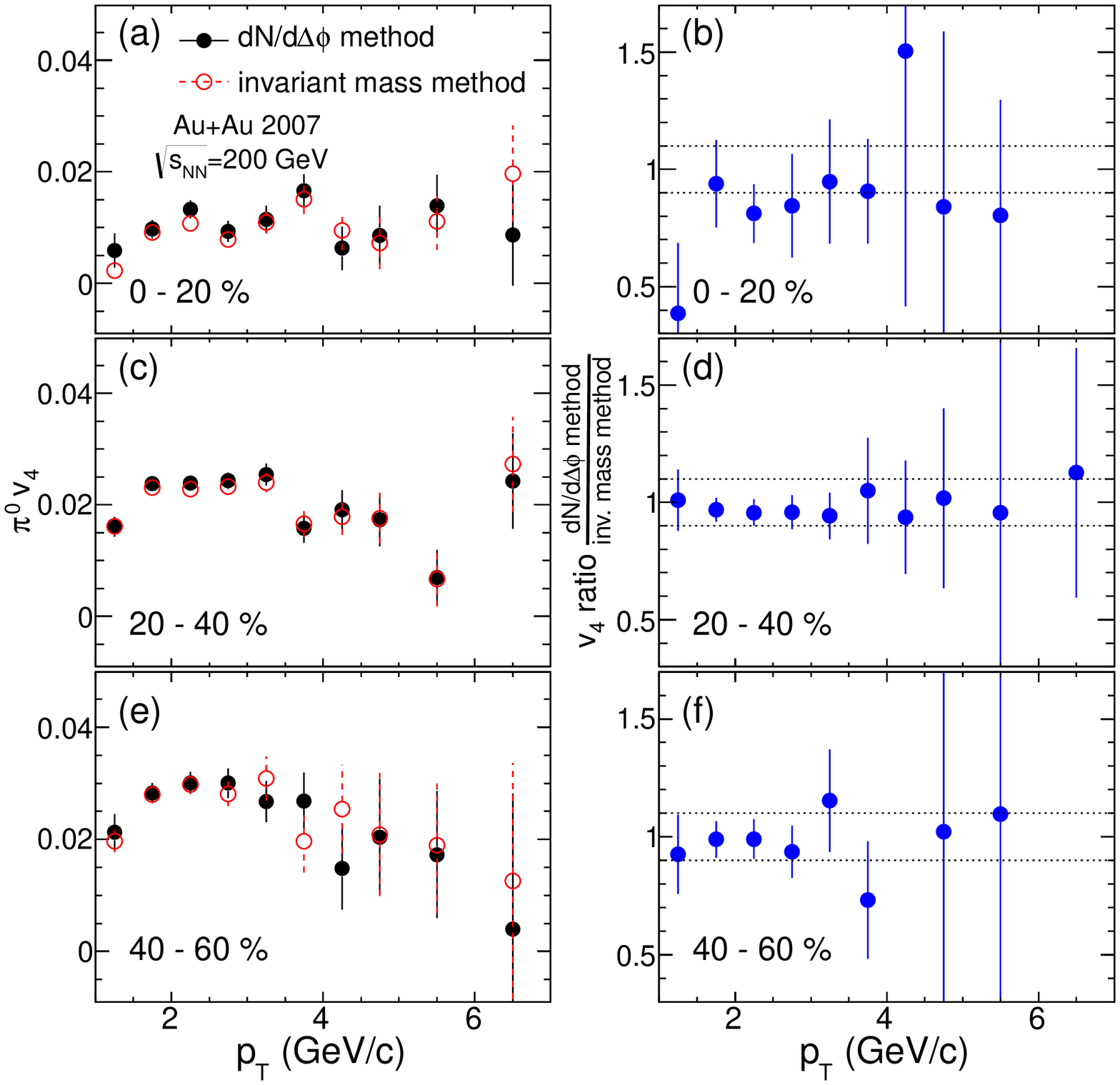}
\end{center}
\caption{\label{fig:pi0m7} (Color online)  Comparison of $\pi^0$ $v_4$ 
for the $dN/d\Delta\phi$ and the invariant mass methods of analysis for 
several centrality selections (left panels). The corresponding ratios 
are shown in the right panels, with the dashed curves indicate a $\pm 
10$\% deviation from unity.}
\end{figure}
\begin{figure}[thb]
\begin{center}
\includegraphics[width=1.0\linewidth]{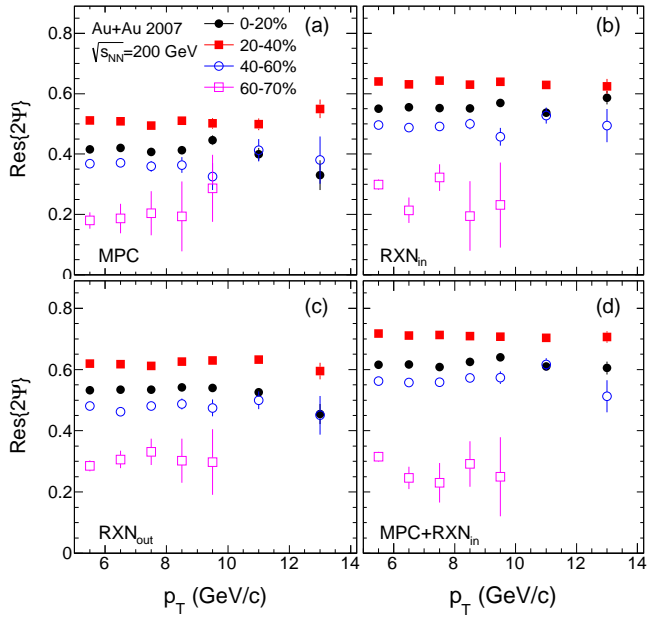}
\end{center}
\caption{\label{fig:pi0m22} (Color online) Event plane resolution 
${{\rm Res}\{2\Psi\}}$ for events containing a high $\pT$ $\pi^0$. 
Results are plotted as a function of $\pT$ of the $\pi^0$ for several EP 
detectors in several centrality ranges.}
\end{figure}
\begin{figure}[thb]
\begin{center}
\includegraphics[width=1.0\linewidth]{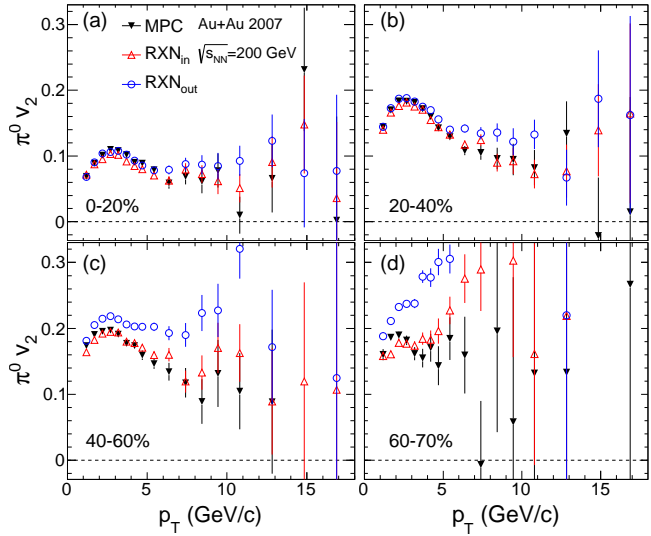}
\end{center}
\caption{\label{fig:pi0m20} (Color online)  Comparison of $v_2$ vs. 
$p_T$ for $\pi^0$ mesons obtained at several centralities with event 
planes determined by detectors in different pseudorapidity ranges.}
\end{figure}
\begin{figure}[thb]
\begin{center}
\includegraphics[width=1.0\linewidth]{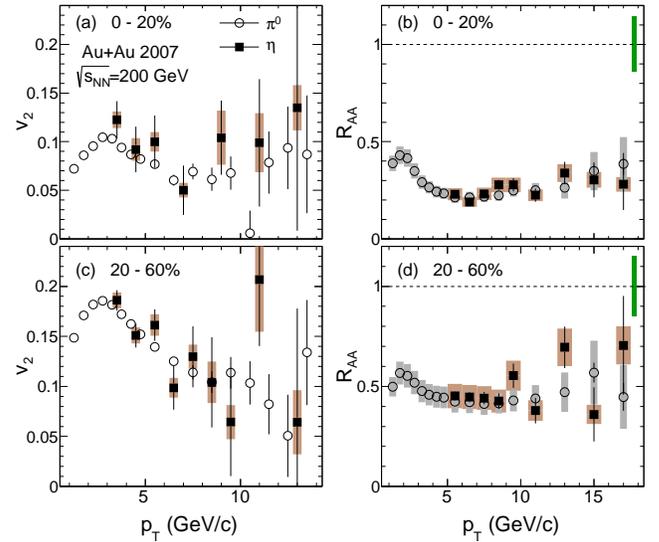}
\end{center}
\caption{\label{fig:rea2c} (Color online) The $v_2$ (left panels) and 
nuclear modification factor $R_{{\rm AA}}$ (right panels) for $\pi^0$ 
and $\eta$ mesons in 0\%--20\% (top panels) and 20\%--60\% (bottom 
panels) centrality ranges. The $v_2$ results for $\pi^0$ mesons are 
identical to what was published in Ref.~\cite{Adare:2010sp}. The 
$R_{{\rm AA}}$ data are taken from Ref.~\cite{Adare:2008qa} ($\pi^0$, 
$\pT<5$ GeV/$c$), Ref.~\cite{Adare:2012wg} ($\pi^0$, $\pT>5$ GeV/$c$) 
and Ref.~\cite{Adare:2010dc} (eta meson). The uncertainties of 
$R_{{\rm AA}}$ associated with $N_{\rm coll}$ and normalization are 
common between $\pi^0$ and $\eta$ mesons, and are shown as shaded boxes 
around unity.}
\end{figure}
\begin{figure}[th]
\begin{center}
\includegraphics[width=1.0\linewidth]{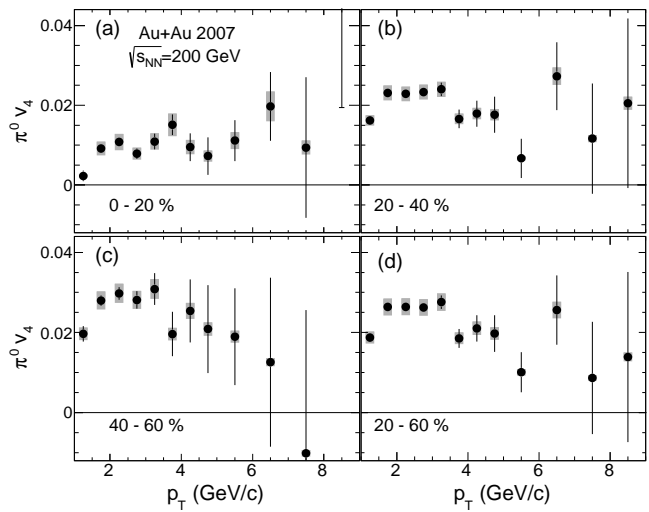}
\end{center}
\caption{\label{fig:rea6} The $v_4$ coefficient vs. $\pT$ for $\pi^0$
mesons for the indicated centrality selections. The shaded 
boxes represent the systematic uncertainties.}
\end{figure} %
\begin{figure}[ht] %
\begin{center}
\includegraphics[width=1.0\linewidth]{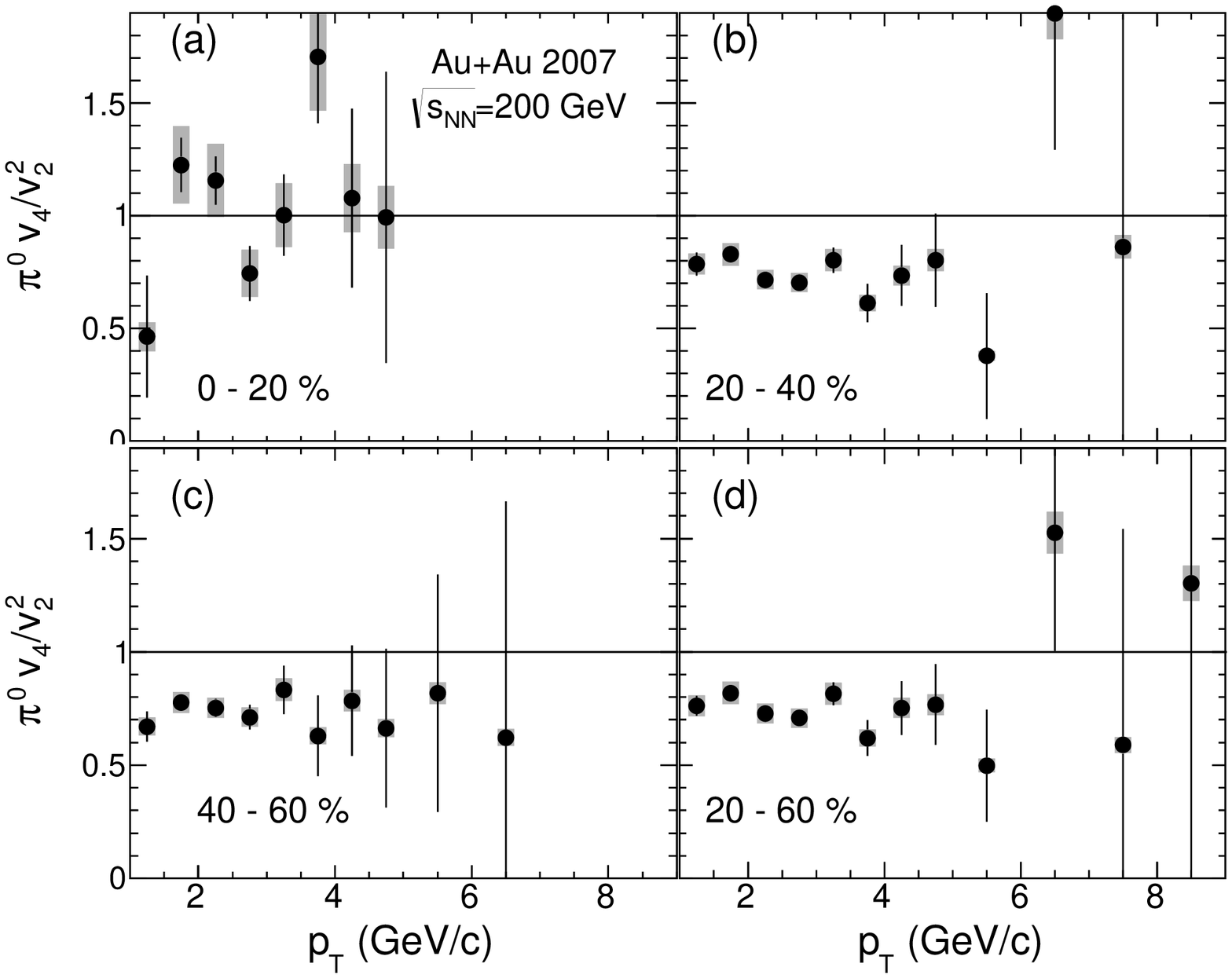}
\end{center}
\caption{\label{fig:rea7} The ratio $v_4/v_2^2$ vs. $\pT$ for $\pi^0$
mesons for the indicated centrality selections. The shaded boxes 
represent the systematic uncertainties.}
\end{figure}

Figures~\ref{fig:pi0m6} and \ref{fig:pi0m7} compare the $v_2$ and $v_4$ 
values obtained from the two analysis methods. The $v_2$ values agree 
within the statistical uncertainties; the systematic deviation is less 
than 3\%. The $v_4$ from the $dN/d\Delta\phi$ method is systematically 
larger by about 5\%--10\%. This is due to small residual backgrounds in 
the integration window in the $dN/d\Delta\phi$ method. This background 
also leads to larger statistical uncertainty for the $dN/d\Delta\phi$ 
method, because it is evaluated separately for each angular bin, while 
in reality this background is correlated between these angular bins. The 
$dN/d\Delta\phi$ method also has an extra source of systematic 
uncertainty arising from the number of $\Delta\phi$ bins, which can 
become significant for $v_4$. Consequently, the invariant mass method is 
used to generate the $v_4$ and $v_4/v_2^2$ results.  As a cross-check, 
the $\pi^0$ $v_2$ and $v_4$ results are also compared and found to be 
consistent with the results for identified charged pions from 
Ref.~\cite{Adare:2010ux}.

\subsubsection{Evaluating the jet bias}

This measurement assumes that the EP determination is not strongly 
biased by the selection of $\pi^0$ and $\eta$ mesons at midrapidity. At 
high $\pt$, such a bias could stem from dijets associated with the 
$\pi^0$ or $\eta$ mesons. The pseudorapidity spread of particles from 
jets containing the $\pi^0$ or $\eta$ meson is typically much smaller 
than the pseudorapidity gap between the EP and the EMCal. However, the 
large rapidity swing of the away-side jet could bring jet-associated 
particles into the detectors used to calculate the EP, leading to a 
potential bias of the $v_{2}$ and $v_4$ values.

In earlier studies of high momentum particles, PHENIX has estimated the 
magnitude of this bias by embedding {\sc pythia} dijet events into 
HIJING events modulated with the experimentally measured $v_2$ 
signal~\cite{Adare:2008ae,Jia:2006sb}. The away-side jet was found to 
bias the EP determination and hence, the $v_{2k}$ signal at high $\pT$, 
depended on the pseudorapidity gap. In general the bias is expected to 
decrease with increasing pseudorapidity gap, and should be smallest for 
EP determined by the MPC pseudorapidity range. In this analysis, our 
ability to measure the EP in different pseudorapidity ranges allows for 
a data-driven quantification of the pseudorapidity dependence of the jet 
bias, as discussed below.

Figure~\ref{fig:pi0m22} shows the EP resolution of various detectors for 
events containing a high $\pT$ $\pi^0$ ($\pT>5$ GeV/$c$). No systematic 
$\pT$ dependence is observed for the selections studied. 
Figure~\ref{fig:pi0m20} compares the $v_2(\pT)$ values for $\pi^0$, 
obtained with event planes determined in the MPC ($3<|\eta|<4$), 
RXN$_{\rm in}$ ($1.5<|\eta|<2.8$) and RXN$_{\rm out}$ ($1<|\eta|<1.5$). 
At low $\pT$ ($\alt5$ GeV/$c$), the $v_2$ values obtained from these 
event planes are comparable in more central collisions. For higher 
$\pT$, they deviate from each other, with larger $v_2$ for RXN$_{\rm 
out}$ and smaller $v_2$ for the MPC. For peripheral collisions, the 
values obtained with RXN$_{\rm out}$ are significantly higher over the 
full $\pT$ range. These trends are consistent with the presence of a 
dijet bias, which grows as the pseudorapidity gap between the EP and the 
$\pi^0$ or $\eta$ meson is reduced. For pseudorapidity gaps $\lesssim 
1$--1.5, the dijets bias the EP angle towards the direction of the 
$\pi^0$, which results in a larger $v_2$ value. Apparently, this bias 
does not affect the resolution corrections. In summary, for the 
centrality range used in this analysis (0\%--60\%), the use of 
MPC+RXN$_{\rm in}$ is sufficient to suppress the effects of this jet 
bias to within the statistical uncertainty of the measurement (see also 
Ref.~\cite{Adare:2010sp}).

\subsubsection{Systematic uncertainties}

The primary results of this analysis are obtained with the MPC+RXN$_{\rm 
in}$ event plane. The uncertainties in the resolution factors for this 
event plane are obtained by comparing the values obtained for the 2SE and 
the 3SE methods. They are estimated to be 8\% (12\%) in central collisions 
and 4\% (6\%) in midcentral collisions for ${\rm Res}\{2\Psi\}$ (${\rm 
Res}\{4\Psi\}$). These uncertainties allow all points to move up and down 
by the same multiplicative factor.

The systematic uncertainties for $v_{2}^{{\rm obs}}$ and $v_{4}^{{\rm 
obs}}$ are estimated by varying the identification cuts for $\pi^0$ and 
$\eta$ mesons, the parameterization of the residual background and the 
peak integration window in the $\mgg$ distributions. These uncertainties 
are correlated in $\pT$ and are added in quadrature to give the total 
systematic uncertainties. For the $\pi^0$ analysis, these uncertainties 
are estimated to be 10\% (15\%) in central collisions and $3\%$ (5\%) in 
midcentral collisions for $v_{2}^{{\rm obs}}$ ($v_{4}^{{\rm obs}}$). For 
the $\eta$ meson analysis, the uncertainties are significantly larger 
primarily because of the lower signal-to-background ratio. These 
uncertainties are estimated to be 15\% in central collisions and 
$\sim10$\% at other centralities.

\section{Results and Discussion}

The primary results of this analysis are obtained with MPC+RXN$_{\rm 
in}$ event plane. The left panels of Fig.~\ref{fig:rea2c} compare the 
$v_2$ for $\pi^0$ and $\eta$ mesons for $\pT\agt4$ GeV/$c$, both 
obtained with the $dN/d\Delta\phi$ method for two centrality ranges. 
Within uncertainties, the magnitude of the $v_2$ values for both 
particle species are the same for the measured $\pT$ range. This 
agreement indicates that the differences between their masses 
[$m_{\eta}=0.548$ GeV, $m_{\pi^0}=0.135$ GeV] and quark content 
[$(u\bar{u}-d\bar{d})/\sqrt{2}$ for $\pi^0$ and 
$(u\bar{u}+d\bar{d}-2s\bar{s})/\sqrt{6}$ for $\eta$ in the SU(3) limit] 
do not lead to appreciable differences in the $\pi^0$- and $\eta$-meson 
$v_2$ values for the centrality ranges studied. A clear decrease of 
$v_2$ with $p_T$ is also evident, especially for the 20\%--60\% 
centrality selection. These patterns complement our earlier suppression 
measurements~\cite{Adare:2012wg,Adare:2010dc} (reproduced in the right 
panel of Fig.~\ref{fig:rea2c}), which show the same suppression patterns 
for $\pi^0$ and $\eta$ mesons. These results are in agreement with the 
expectations for in-medium energy loss of parent partons prior to their 
fragmentation into hadrons in the $\pT$ region where jet quenching is 
expected to be dominant mechanism ($\pT\agt4$--5 GeV/$c$).

The transition from anisotropy driven by hydrodynamic flow to anisotropy 
driven by jet quenching can be probed by the ratio $v_4/v_2^2$. Perfect 
fluid hydrodynamics predicts a value of 0.5 for this 
ratio~\cite{Borghini:2005kd}. However, geometrical fluctuations and 
other dynamical fluctuations, as well as viscous damping, can 
significantly increase the magnitude of this ratio, especially in 
central collisions~\cite{Gombeaud:2009ye,Lacey:2011ug}. Furthermore, the 
directions that maximize collective flow and jet quenching may not be 
the same~\cite{Jia:2012ez,Zhang:2012ha}. Hence, this ratio could change 
in the $\pT$ region where jet quenching begins to dominate.

The results of $v_4(\pT)$ and the $v_4/v_2^2$ ratio for $\pi^0$ in several 
centrality ranges are shown in Fig.~\ref{fig:rea6} and 
Fig.~\ref{fig:rea7}, respectively. Results are also combined into a wide 
centrality range (20\%--60\%) for better statistical precision. 
Figure~\ref{fig:rea6} shows that significant $v_4$ values are observed 
even for $\pT>5$ GeV/$c$. The $v_4/v_2^2$ ratios shown in 
Fig.~\ref{fig:rea7} are approximately independent of $\pT$, with values of 
$\sim$1.0 and $\sim$0.8 for the 0\%--20\% and the 20\%--40\% and 40\%--60\% 
centrality selections respectively. This pattern at low $\pT$ 
($\pT\lesssim 5$ GeV/$c$) is consistent with our prior observations of 
this ratio for inclusive charged hadron 
measurements~\cite{Adare:2010ux}. On the other hand, possible variations 
of this ratio at higher $\pT$ could be masked, owing to the limited 
statistics of this measurement. The deviation of the $v_4/v_2^2$ value 
from the expectation of ideal hydrodynamics and the increase of this 
ratio from midcentral to more central collisions may reflect the 
combined effects of fluctuations in the initial geometry and finite 
viscosity in the evolving medium~\cite{Gombeaud:2009ye}.

\section{Summary and Conclusions}

PHENIX has measured the azimuthal anisotropy for $\pi^0$ and $\eta$ 
mesons in Au$+$Au collisions at $\sqrt{s_{\rm NN}}=200$ GeV. The 
anisotropy coefficients $v_2$ and $v_4$ are measured with event planes 
determined in forward detectors, which enable a minimum pseudorapidity 
gap of $1.2$ units between the event plane and the $\pi^0$ or $\eta$ 
mesons. This pseudorapidity gap is found to greatly reduce 
auto-correlation biases due to dijets over the 0\%--60\% centrality 
range. The magnitude of the $v_2$ values extracted for $\pi^0$ and 
$\eta$ mesons, over the common $\pT$ range of 3--14 GeV/$c$, are 
observed to be similar, suggesting in-medium energy loss of parent 
partons prior to their fragmentation into hadrons. The $v_4$ values for 
$\pi^0$ mesons are found to be significantly above zero for the measured 
$\pT$ range of 1--9 GeV/$c$. The $v_4/v_2^2$ ratios are independent of 
$\pT$ with a magnitude between ${\sim}0.8$ and ${\sim}1.0$ depending on 
the centrality range, which may reflect the combined effects of 
fluctuations in the initial collision geometry and finite viscosity in 
the evolving medium.



\section*{Acknowledgments}   

We thank the staff of the Collider-Accelerator and Physics
Departments at Brookhaven National Laboratory and the staff of
the other PHENIX participating institutions for their vital
contributions.  We acknowledge support from the 
Office of Nuclear Physics in the
Office of Science of the Department of Energy, the
National Science Foundation, Abilene Christian University
Research Council, Research Foundation of SUNY, and Dean of the
College of Arts and Sciences, Vanderbilt University (U.S.A),
Ministry of Education, Culture, Sports, Science, and Technology
and the Japan Society for the Promotion of Science (Japan),
Conselho Nacional de Desenvolvimento Cient\'{\i}fico e
Tecnol{\'o}gico and Funda\c c{\~a}o de Amparo {\`a} Pesquisa do
Estado de S{\~a}o Paulo (Brazil),
Natural Science Foundation of China (P.~R.~China),
Ministry of Education, Youth and Sports (Czech Republic),
Centre National de la Recherche Scientifique, Commissariat
{\`a} l'{\'E}nergie Atomique, and Institut National de Physique
Nucl{\'e}aire et de Physique des Particules (France),
Bundesministerium f\"ur Bildung und Forschung, Deutscher
Akademischer Austausch Dienst, and Alexander von Humboldt Stiftung (Germany),
Hungarian National Science Fund, OTKA (Hungary), 
Department of Atomic Energy and Department of Science and Technology (India), 
Israel Science Foundation (Israel), 
National Research Foundation and WCU program of the 
Ministry Education Science and Technology (Korea),
Physics Department, Lahore University of Management Sciences (Pakistan),
Ministry of Education and Science, Russian Academy of Sciences,
Federal Agency of Atomic Energy (Russia),
VR and Wallenberg Foundation (Sweden), 
the U.S. Civilian Research and Development Foundation for the
Independent States of the Former Soviet Union, 
the US-Hungarian Fulbright Foundation for Educational Exchange,
and the US-Israel Binational Science Foundation.


\bibliographystyle{apsrev4-1}

\end{document}